\documentclass[12pt]{article}

\usepackage{newtxtext,newtxmath}

\usepackage{graphicx}

\usepackage[letterpaper,margin=1in]{geometry}

\renewenvironment{abstract}
	{\quotation}
	{\endquotation}

\date{}

\makeatletter
\renewcommand{\fnum@figure}{\textbf{Figure \thefigure}}
\renewcommand{\fnum@table}{\textbf{Table \thetable}}
\makeatother

\usepackage{scicite}

\usepackage{url}

\def\scititle{
	Periodic extreme rainfall in a warmer climate due to stronger convectively-coupled waves
}
\title{\bfseries \boldmath \scititle}

\author{
	Heng~Quan$^{1,2\ast}$,
	Yi~Zhang$^{3}$,
	Guy~Dagan$^{4}$,
    Stephan~Fueglistaler$^{1,2}$\and
	\small$^{1}$Department of Geosciences, Princeton University; Princeton; NJ; USA.\and
    \small$^{2}$Program in Atmospheric and Oceanic Sciences, Princeton University; Princeton; NJ; USA.\and
	\small$^{3}$Courant Institute of Mathematical Sciences, New York University; New York; NY; USA.\and
    \small$^{4}$Institute of Earth Sciences, The Hebrew University of Jerusalem; Jerusalem; Israel.\and
	\small$^\ast$Corresponding author(s). Email(s): hengquan@princeton.edu\and
}

\begin{document} 

\maketitle

\begin{abstract} \bfseries \boldmath
Tropical regions may experience periodic extreme precipitation and suffer from associated periodic deluges in a warmer climate. Recent studies conducted small-domain (around 100\,km $\times$ 100\,km) atmospheric model simulations and found that precipitation transitions from a steady state to a periodic oscillation state in a hothouse climate when the sea surface temperature reaches 320-325\,K. Here we conduct global-scale atmospheric model simulations with different complexity, and we find that tropical precipitation in convective regions already transitions to a $\mathcal{O}(10\,\mathrm{day})$ periodic oscillation state with a $\mathcal{O}(100\,\mathrm{mm\,day^{-1}})$ amplitude at 305-310\,K. This temperature is substantially lower than previously reported, and within reach in a century under a high carbon emission scenario. We attribute the onset of the periodic extreme precipitation to the intensification of convectively-coupled waves, which occurs at temperatures much lower than the radiative mechanism responsible for the transition around 320-325\,K identified before.
\end{abstract}

\noindent

Extreme precipitation events can cause severe economic loss and damage to natural systems \cite{ipccar6}, and they are projected to be more intense and frequent under global warming \cite{min2011,lehmann2015,guerreiro2018}. Tropical regions, where the baseline precipitation is already high, are especially vulnerable to more severe extreme precipitation events in the future. In addition, projections of human population predict a continuing growth in countries in the
tropics ($20^\circ$S to $20^\circ$N) \cite{legg2024}. Therefore, there is a pressing need to better understand the features and mechanisms of tropical extreme precipitation in a warmer climate down to regional scales. 

Recent studies have found that tropical precipitation transitions to a periodic oscillation state in hothouse climates \cite{seeley2021,song2024,dagan2023,spaulding2024,yang2024}. These studies performed idealized small-domain (around 100\,km $\times$ 100\,km) simulations that explicitly resolve convection, and found that precipitation only occurs in short outbursts (around 100\,$\mathrm{mm\,day^{-1}}$) lasting a few hours separated by multi-day dry spells when the sea surface temperature (SST) reaches 320-325\,K. This periodic extreme precipitation state would leave soils saturated, rivers swollen, and communities unable to recover before the next deluge strikes, resulting in a total damage far exceeding that of isolated extreme precipitation events of similar intensity. However, the critical SST is 20-25\,K warmer than tropical SSTs in present climate. Such a hothouse climate state can be found in Earth's distant past with extremely high carbon dioxide levels \cite{sleep2010,charnay2017,pierrehumbert2011} and potentially in the distant future with much larger solar insolation \cite{goldblatt2012}, but is unrealistic in the near future \cite{ipccar6}.

Here, we indicate that global-scale simulations show a similar transition to periodic precipitation in a climate only 5-10\,K warmer than present, which is much lower than expected by previous work (320-325\,K) based on small-domain simulations. Model projections show that for high-emission scenarios, the surface temperature may increase by 5-10\,K in a century \cite{ipccar6}. We find that precipitation in tropical convective regions features a $\mathcal{O}(10\,\mathrm{day})$ periodic oscillation with a $\mathcal{O}(100\,\mathrm{mm\,day^{-1}})$ amplitude when the local SST reaches 305-310\,K, resulting in severe periodic deluges. We attribute the transition to such a periodic precipitation state to the intensification of large-scale convectively-coupled waves, which was not considered by previous work based on small-domain simulations. Our results demonstrate that the coupling between tropical waves and convection can significantly lower the temperature at which a transition to periodic extreme precipitation may be observed.

\subsection*{Global climate model projections}

Figure \ref{fig:AM4_precip_series} shows the response of tropical precipitation to global warming from global atmospheric model simulations with GFDL-AM4 \cite{zhao2018a,zhao2018b} forced by SSTs in present climate and warmer climates. Figure \ref{fig:AM4_precip_series}a-b shows that the climatological mean precipitation becomes stronger almost everywhere between $20^\circ$S to $20^\circ$N in a warmer climate, consistent with the ``wet-get-wetter" argument by \cite{held2006}. In parallel, a transition from steady precipitation to periodic precipitation down to regional scales is evident in Fig.\,\ref{fig:AM4_precip_series}c-f, which shows the time series of 3-hourly precipitation in three climate states in four selected convective regions near the equator (red boxes in Figure \ref{fig:AM4_precip_series}b). Precipitation features small fluctuations about a steady mean state in present climate (blue curves), but transitions to a $\mathcal{O}(10\,\mathrm{day})$ periodic oscillation with a $\mathcal{O}(100\,\mathrm{mm\,day^{-1}})$ amplitude in a 5-10\,K warmer climate (black and red curves). While the increase in mean precipitation may be expected to project onto the amplitude of the oscillation, the power spectrum of normalized precipitation by its regional mean (Fig.\,\ref{fig:AM4_precip_series}g-j, see Materials and Methods) also shows that the spectral power of the $\mathcal{O}(10\,\mathrm{day})$ oscillation amplifies by a factor of 2-4 in a 5-10\,K warmer climate.

Targeted small-domain cloud-resolving model simulations (Fig.\ref{fig:SAM_precip_series}a) confirm the previously reported transition from steady precipitation to periodic precipitation with a $\mathcal{O}(100\,\mathrm{mm\,day^{-1}})$ amplitude when the SST reaches 320-325\,K \cite{seeley2021,song2024,dagan2023,spaulding2024,yang2024}. Previous work explained this transition to result from a net lower-tropospheric radiative heating due to the closing of the water vapor infrared window regions \cite{seeley2021}, while following studies found that the lower-tropospheric radiative heating is not necessary. A large vertical gradient in radiative cooling rate - close to zero in lower troposphere but strong in upper-troposphere - can also help to build up convective instability in the inhibition phase and trigger periodic convection \cite{dagan2023,song2024}. However, the transition to periodic precipitation in global climate model simulations occurs at much lower SSTs around 305-310\,K, at which the vertical gradient of the time-average radiative cooling rate does not increase significantly in all selected regions (Fig.\ref{fig:AM4_QRAD_tmean}). Moreover, the radiative cooling rate does not always show a large vertical gradient in inhibition phases (Fig.\ref{fig:AM4_QRAD_series}). Therefore, the transition to periodic precipitation in global climate model simulations is a consequence of a different process from previously reported.

In the following, we argue that the transition to periodic precipitation in global climate model simulations is instead related to the intensification of large-scale convectively-coupled waves, which is not captured by small-domain simulations in previous work. The tropical atmosphere sustains a group of convectively coupled wave modes \cite{Matsuno1966,Gill1980} that control a substantial fraction of tropical precipitation variability \cite{wheeler1999,kiladis2009}. The global climate model simulations shown in Fig.\,\ref{fig:AM4_precip_series} exhibit an intensification of the eastward-propagating convectively-coupled Kelvin waves under global warming, both in the Hovmoller diagrams (i.e., the time-longitude plot, see Fig.\ref{fig:AM4_precip_Hovmoller}) and the wave spectrum diagrams (Fig.\ref{fig:AM4_EQwaves_symmetric}) of equatorial precipitation. This result from GFDL-AM4 is consistent with the intensification of Kelvin waves under global warming seen in almost all CMIP6 global climate models \cite{bartana2023}. The intensification of a $\mathcal{O}(10\,\mathrm{day})$ wave signal (Fig.\ref{fig:AM4_EQwaves_symmetric}) is consistent with the periodic oscillation of precipitation in different regions in a 5-10\,K warmer climate (Fig.\ref{fig:AM4_precip_series}c-f), implying that stronger convectively-coupled waves are responsible for the transition to periodic precipitation under global warming. However, global climate model simulations have biases representing convectively-coupled waves due to coarse resolution ($\mathcal{O}(100\,\mathrm{km})$) and convective parameterization \cite{dias2018,weber2021,ji2025}. Therefore, we turn to more idealized large-scale cloud-resolving simulations with a much higher resolution (2\,km) to demonstrate that convectively-coupled waves will intensify and trigger periodic precipitation in a warmer climate.

\subsection*{Cloud-resolving model simulations}

We use the cloud-resolving model SAM \cite{khairoutdinov2003} to study convectively-coupled waves. We conduct mock Walker simulations \cite{kuang2012} in a large domain ($12288\,\mathrm{km} \times 128\,\mathrm{km}$) with a horizontal resolution of 2\,km (see Materials and Methods). These simulations have a linearly-varying SST along the long, planetary-scale dimension, which mimics the east-west SST gradient across the equatorial Pacific and results in an overturning circulation similar to the Walker circulation in the tropical Pacific. We ignore the planetary rotation since it is weak close to the equator, so the mock Walker simulations only sustain convectively-coupled gravity waves, which differs from the convectively-coupled Kelvin waves in global climate model simulations. However, studying the convectively-coupled gravity waves in mock Walker simulations proves useful to understand the convectively-coupled Kelvin waves in global climate model simulations. As we will show below, they have the same propagation behavior, respond similarly to global warming, and both trigger periodic precipitation in a warmer climate because they have the same dispersion relation $\omega = ck$.

Figure \ref{fig:SAM_precip_series}c shows the transition to periodic precipitation under global warming in mock Walker simulations. The 300\,K, 305\,K and 310\,K mock Walker simulations are comparable to the present, +5\,K and +10\,K global climate model simulations (Fig.\ref{fig:AM4_precip_series}c-f), respectively. Convective regions in both models exhibit a transition to a $\mathcal{O}(10\,\mathrm{day})$ periodic precipitation with a $\mathcal{O}(100\,\mathrm{mm\,day^{-1}})$ amplitude in a 5-10\,K warmer climate, whereas the small-domain simulations (Fig.\ref{fig:SAM_precip_series}a) show the transition at the previously reported SST of 320-325\,K. Simulations with identical setup but a prescribed uniform radiative cooling rate ($Q_{\mathrm{rad}} = -1.5\,\mathrm{K\,day^{-1}}$) in the troposphere show the same transition  (Fig.\ref{fig:SAM_precip_series_mock_Walker_uniform_QRAD}), confirming that changes in radiative cooling are not the cause of the transition. Instead, the intensification of large-scale convectively-coupled gravity waves is evident when looking at the Hovmoller diagrams of precipitation (Fig.\ref{fig:SAM_mock_Walker_Hovmoller}). Consistent with previous studies \cite{grabowski2000,dagan2023}, we find that the periods of precipitation in the 305\,K and 310\,K mock Walker simulations are consistent with the corresponding periods of gravity waves (Fig.\ref{fig:SAM_mock_Walker_Hovmoller}d-e). Moreover, the period of precipitation is proportional to the length of the domain in additional simulations with the same SST (i.e., same phase speed of gravity waves) but varying domain sizes (Fig.\ref{fig:SAM_precip_series_mock_Walker_perturbL}), further confirming the coupling between convection and gravity waves. Both the global climate model simulations and the mock Walker simulations imply that convectively-coupled waves play a central role in the transition to periodic precipitation.

To demonstrate that convectively-coupled waves will intensify and trigger periodic precipitation in a warmer climate, we isolate its effects by running a new suite of cloud-resolving simulations, where we couple the small-domain simulations to large-scale gravity waves following Ref. \cite{kuang2008DGW} (see Materials and Methods). For simplicity we assume the large-scale gravity waves have a single horizontal wavelength of 8000\,km, which is similar to the wavelength of Kelvin waves in global climate model simulations (Fig.\ref{fig:AM4_EQwaves_symmetric}) and comparable to the long dimension of the mock Walker simulations (12288\,km), but much larger than the size of the small domain (128\,km). As a result, the atmospheric motion in the entire small domain is largely synchronized by gravity waves. Figure \ref{fig:SAM_precip_series}b shows that the coupled simulations also exhibit a transition to a $\mathcal{O}(10\,\mathrm{day})$ periodic precipitation with a $\mathcal{O}(100\,\mathrm{mm\,day^{-1}})$ amplitude when the SST reaches 305-310\,K, consistent with the mock Walker simulations (Fig.\ref{fig:SAM_precip_series}c) and global climate model simulations (Fig.\ref{fig:AM4_precip_series}c-f). The comparison between Figure \ref{fig:SAM_precip_series}a and Figure \ref{fig:SAM_precip_series}b indicates that large-scale convectively-coupled waves alone can significantly lower the temperature for the transition to periodic precipitation. Therefore, we consider the coupled simulations in Figure \ref{fig:SAM_precip_series}b as the minimum recipe capturing the transition to periodic precipitation in a warmer climate. In the following, we will analyze the coupled simulations to diagnose the mechanism behind the transition.

\subsection*{Mechanism of the transition}

Figure \ref{fig:SAM_WTG_mechanism} demonstrates that the periodic precipitation is due to the self-reinforcement of convectively-coupled waves. In a background radiative convective equilibrium (RCE) state, the potential temperature $\theta$ increases with height (Fig.\ref{fig:SAM_WTG_mechanism}a) while the specific humidity $q$ decreases with height (Fig.\ref{fig:SAM_WTG_mechanism}b) in the troposphere. When a passing wave causes a domain-wise ascending motion and triggers convection in the lower troposphere, the vertical advection brings air with smaller $\theta$ and larger $q$ upwards, which cools and moistens the lower troposphere (Fig.\ref{fig:SAM_WTG_mechanism}c and equation \ref{equ:T_tendency}-\ref{equ:q_tendency} in Materials and Methods). A colder temperature and a larger specific humidity both increase relative humidity and inhibit the evaporation of cloud condensates, which allows the convection to reach higher and results in more intense precipitation \cite{kuang2008instability,ahmed2020,weber2021}. By contrast, the convection is suppressed in the opposite phase of gravity waves with domain-wise descending motion in the lower troposphere, resulting in little precipitation. Such a positive feedback deviates the strength of convection away from the steady state, and manifests itself as the periodic oscillation in the time series of precipitation.

With a warmer SST, the potential temperature increases more in the upper troposphere, while the specific humidity increases more in the lower troposphere, resulting in larger vertical gradients of $\theta$ and $q$ in the RCE state (Fig.\ref{fig:SAM_WTG_mechanism}a-b). Therefore, the vertical advection more effectively cools and moistens the lower troposphere in the ascending phase of gravity waves (and vice versa in the descending phase), which amplifies the positive feedback shown in Figure \ref{fig:SAM_WTG_mechanism}c. When the SST is at 290-300\,K, the positive feedback of convectively-coupled waves is not strong enough to trigger instability due to small vertical gradients of $\theta$ and $q$, so the precipitation is in a steady state. When the SST reaches 305-310\,K, the positive feedback becomes strong enough due to large vertical gradients of $\theta$ and $q$, so that the convectively-coupled waves become unstable and the precipitation transitions to a periodic oscillation state (Fig.\ref{fig:SAM_precip_series}b). The mechanism sketched in Figure \ref{fig:SAM_WTG_mechanism}c would produce a monotonic increase in the amplitude of convectively-coupled waves under global warming, and future work may address the question what process leads to the non-linear, threshold-like transition in Figure \ref{fig:SAM_precip_series}b. 

To further verify the proposed mechanism, we run another group of simulations to isolate the effects of the vertical $\theta$ and $q$ gradients. We take the 300\,K simulation in Figure \ref{fig:SAM_precip_series}b, which features steady precipitation, as the base simulation. We increase the vertical gradients of $\theta$ or $q$ approximately to the corresponding values in the 310\,K simulation when calculating the wave-induced advective tendencies (Fig.\ref{fig:SAM_WTG_mechanism}c and equation \ref{equ:T_tendency}-\ref{equ:q_tendency} in Materials and Methods), while fixing everything else including the SST to 300\,K. We find that a larger vertical $\theta$ and $q$ gradient alone can both trigger instability and cause the transition to periodic precipitation, although a larger $q$ gradient is more effective (Fig.\ref{fig:SAM_precip_series_WTG_+Tqgrad}). These results demonstrate that the transition to periodic precipitation in a warmer climate is ultimately the result of larger vertical gradients of $\theta$ and $q$ in the troposphere.

\begin{figure}
    \centering
    \includegraphics[width=\linewidth]{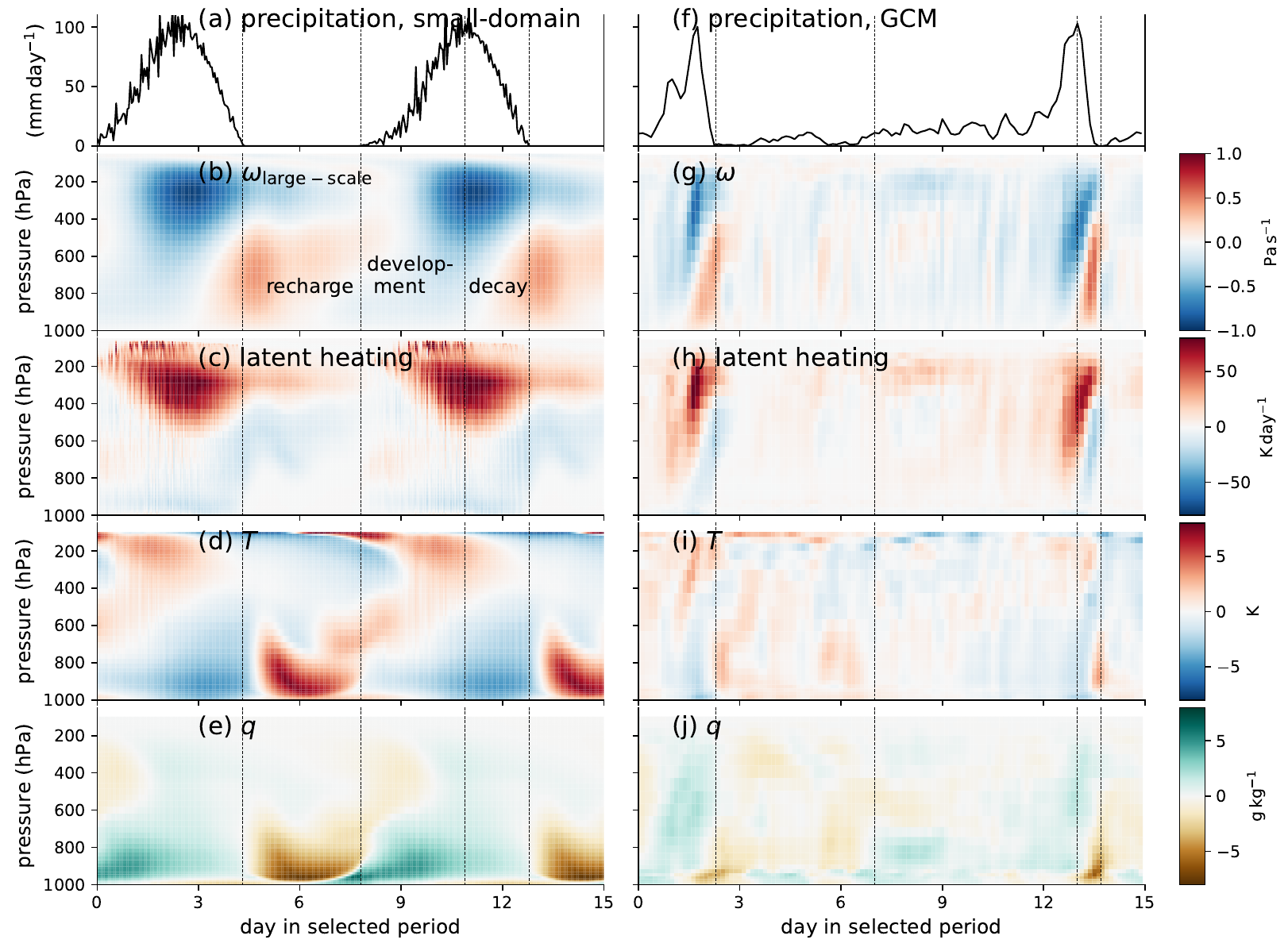}
    \caption{\textbf{The precipitation cycles in a warmer climate.} The left column shows the 15-day time series of domain-mean variables from the 310\,K small-domain cloud-resolving simulation coupled to gravity waves, and the right column shows the time series of the same variables averaged over the selected region in Indonesia from the +10\,K global climate model (GCM) simulation. \textbf{a,e}, Precipitation. \textbf{b,g}, The large-scale vertical velocity induced by gravity waves. \textbf{c,h}, Latent heating rate. \textbf{d,i}, Temperature anomaly relative to the steady state. \textbf{e,j}, Specific humidity anomaly relative to the steady state.}
    \label{fig:convection_cycle_comparison}
\end{figure}

Figure \ref{fig:convection_cycle_comparison}a-e depicts the convection cycles in the 310\,K small-domain simulation coupled to gravity waves. We divide each cycle into three phases: recharge, development, and decay. The recharge phase is a dry period with little or no precipitation (Fig.\ref{fig:convection_cycle_comparison}a), in which the large-scale passing wave causes a domain-wise descending motion in the lower troposphere (Fig.\ref{fig:convection_cycle_comparison}b). The vertical advection in the recharge phase thus warms (Fig.\ref{fig:convection_cycle_comparison}d) and drys (Fig.\ref{fig:convection_cycle_comparison}e) the lower troposphere, building up convective instability. The development phase features increasing precipitation (Fig.\ref{fig:convection_cycle_comparison}a), in which the unstable vertical stratification starts to trigger convection. The wave-induced domain-wise ascending motion in the development phase (Fig.\ref{fig:convection_cycle_comparison}b) cools (Fig.\ref{fig:convection_cycle_comparison}d) and moistens (Fig.\ref{fig:convection_cycle_comparison}e) the lower troposphere, inhibiting the evaporation of cloud condensates. This allows the convection to reach higher altitudes, resulting in latent heating of the upper troposphere by elevated condensation and latent cooling of the lower troposphere by evaporation of precipitation (Fig.\ref{fig:convection_cycle_comparison}c), which reinforces the temperature anomaly. Such a positive feedback process persists until the upper-tropospheric warming and lower-tropospheric cooling stabilize the vertical stratification and start to suppress convection, when the precipitation reaches the peak. After that, the convection weakens and the precipitation decreases in the decay phase (Fig.\ref{fig:convection_cycle_comparison}a), and the cycle restarts from the recharge phase.

The convection cycles in the mock Walker simulations (Fig.\ref{fig:SAM_mock_Walker_convection_cycle}) are consistent with Figure \ref{fig:convection_cycle_comparison}, confirming that the coupling between convection and large-scale atmospheric dynamics can be minimally represented by the coupling between convection and large-scale gravity waves with a single horizontal wavenumber. The convection cycles in the global climate model simulations (Fig.\ref{fig:convection_cycle_comparison}f-j) differ from Figure \ref{fig:convection_cycle_comparison}a-e in a longer period, a shorter time with intense precipitation, and a noisier vertical velocity structure. These differences are expected since the global climate model sustains a group of waves with different wavenumbers, whereas the cloud-resolving model only sustains gravity waves with a single horizontal wavenumber. Nevertheless, the key positive feedback process in Figure \ref{fig:SAM_WTG_mechanism}c - the lower-tropospheric cooling and moistening in the convection development phase and the subsequent top-heavy latent heating - is evident in both models. The Kelvin waves that are dominant in the global climate model simulations (Fig.\ref{fig:AM4_EQwaves_symmetric}) have the same dispersion relation ($\omega = ck$) as the gravity waves in the cloud-resolving simulations, so they couple to convection in the same way \cite{weber2021}. Therefore, the two waves intensify similarly with increasing SST and trigger periodic convection, resulting in qualitatively consistent convection cycles in two models.

\subsection*{Discussion and Conclusion}

\begin{figure}
    \centering
    \includegraphics[width=\linewidth]{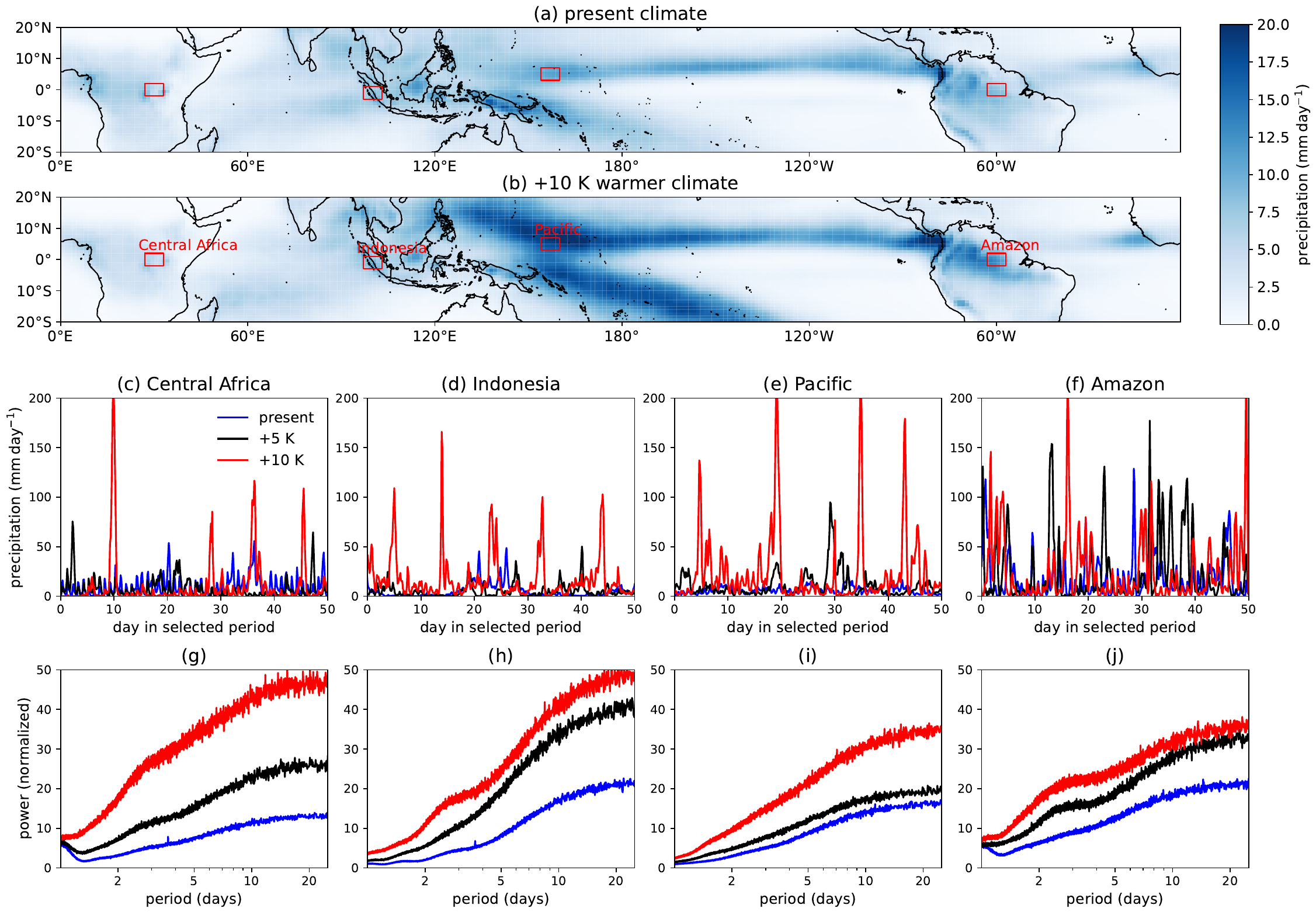}
    \caption{\textbf{Transition to periodic precipitation in global climate model simulations.} \textbf{a},\textbf{b}, Tropical ($20^\circ$S-$20^\circ$N) precipitation in present climate and a +10\,K warmer climate. The results are averages over the last 10 years in 15-year simulations of the atmospheric model GFDL-AM4 (see Materials and Methods). 
    \textbf{c-f}, Regional mean 3-hourly precipitation time series in a selected 50-day period for four regions, in present climate (blue), a +5\,K warmer climate (black) and a +10\,K warmer climate (red). The results are from day 280 to 330 for Central Africa ($3^\circ$S-$3^\circ$N, $26^\circ$E-$32^\circ$E), day 53 to 103 for Indonesia ($4^\circ$S-$2^\circ$N, $95^\circ$E-$101^\circ$E), day 132 to 182 for Pacific ($2^\circ$N-$8^\circ$N, $154^\circ$E-$160^\circ$E) and day 113 to 163 for Amazon ($3^\circ$S-$3^\circ$N, $117^\circ$W-$123^\circ$E), all in the last simulation year.
    \textbf{g-j}, The power spectrums of the 3-hourly precipitation time series (normalized by the mean precipitation) over the last 10 years in all 15-year simulations based on a bootstrap method (see Materials and Methods).}
    \label{fig:AM4_precip_series}
\end{figure}

\begin{figure}
    \centering
    \includegraphics[width=\linewidth]{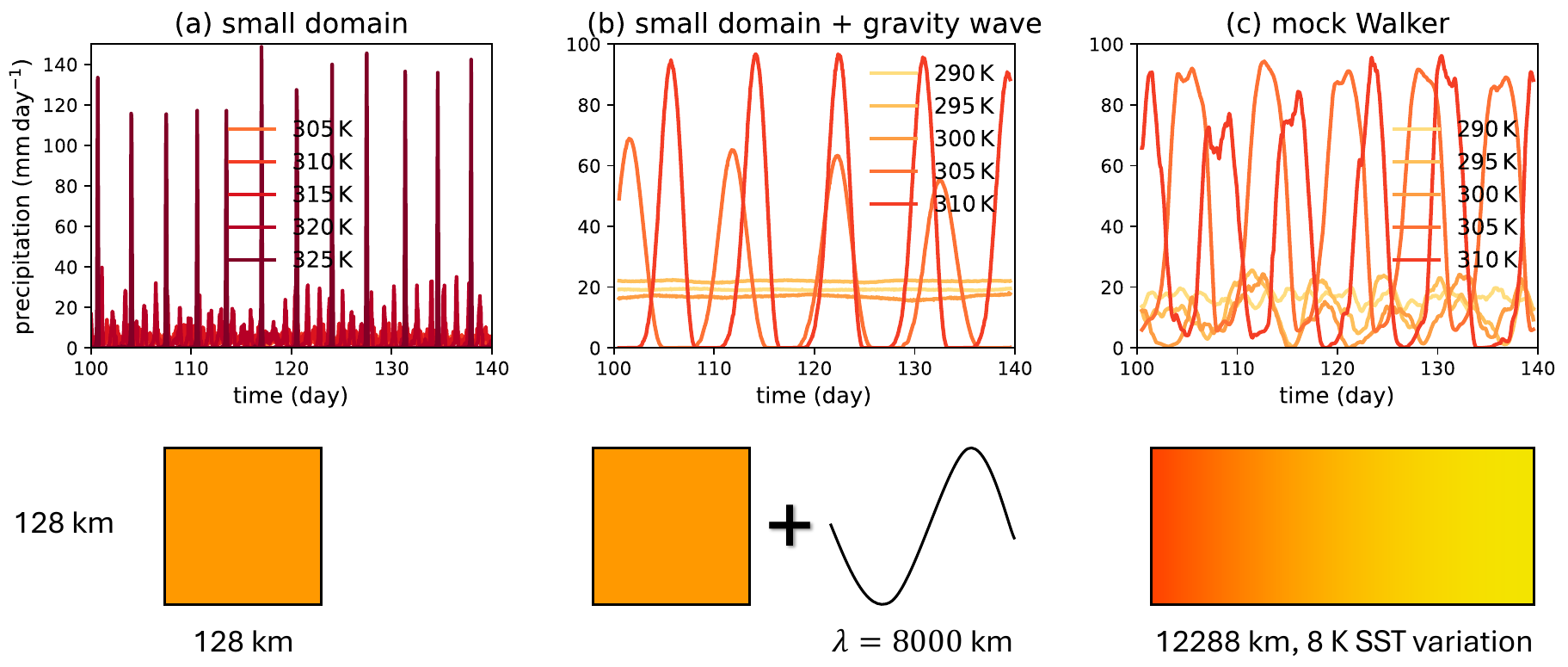}
    \caption{\textbf{Transition to periodic precipitation in cloud-resolving model simulations.} \textbf{a}, 40-day domain-mean precipitation time series in small-domain simulations, with SST ranging from 305\,K to 325\,K with an increment of +5\,K. \textbf{b}, 40-day domain-mean precipitation time series in small-domain simulations coupled to gravity waves, with SST ranging from 290\,K to 310\,K with an increment of +5\,K. \textbf{c}, 40-day precipitation time series averaged over the top 10\% SSTs in 3-D Mock Walker simulations, with the domain-mean SST ranging from 290\,K to 310\,K with an increment of +5\,K.}
    \label{fig:SAM_precip_series}
\end{figure}

\begin{figure}
    \centering
    \includegraphics[width=\linewidth]{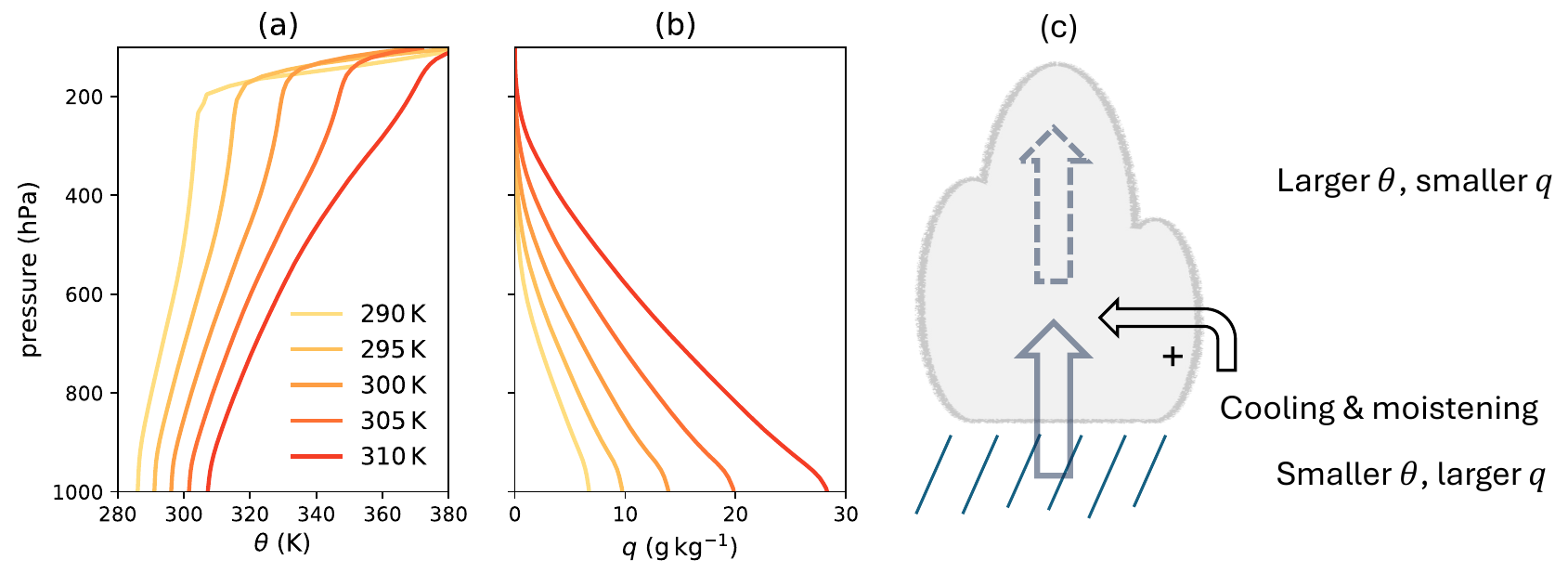}
    \caption{\textbf{The periodic precipitation is due to the self-reinforcement of convectively-coupled waves.} \textbf{a},\textbf{b}, The steady-state (i.e., radiative convective equilibrium) vertical profiles of domain-mean potential temperature $\theta$ and specific humidity $q$ in small-domain cloud-resolving simulations, with SST ranging from 290\,K to 310\,K with an increment of +5\,K. \textbf{c}, A schematic showing how convectively-coupled waves reinforce themselves.}
    \label{fig:SAM_WTG_mechanism}
\end{figure}

A few events similar to the periodic precipitation state described here have been reported in observations. For example, a periodic precipitation event associated with convectively-coupled equatorial Kelvin waves was observed over the equatorial Indian Ocean in early November 2011 during the DYNAMO field campaign \cite{weber2021}. Consistent with our theories, that event is explained by the self-reinforcement of convectively-coupled waves, as depicted in Figure \ref{fig:SAM_WTG_mechanism}c. Our numerical simulations with different complexity all suggest that the periodic precipitation will be more intense and frequent in convective regions in a 5-10\,K warmer climate, which is much lower than previously reported. We attribute the transition to a $\mathcal{O}(10\,\mathrm{day})$ periodic precipitation with a $\mathcal{O}(100\,\mathrm{mm\,day^{-1}})$ amplitude in a 5-10\,K warmer climate to the intensification of convectively-coupled waves, which is ultimately due to larger vertical $\theta$ and $q$ gradients with increasing SST. The transition in idealized small-domain cloud-resolving simulations coupled with gravity waves is abrupt (Fig.\ref{fig:SAM_precip_series}b), while the transition in global climate model simulations is more gradual (Fig.\ref{fig:AM4_precip_series}g-j) due to factors other than convectively-coupled waves. The periodic precipitation state we find is a temporal aggregation of convection, while previous studies documented the spatial aggregation of convection under global warming \cite{muller2012,wing2018,yao2023}. To what extent the temporal aggregation interacts with the spatial aggregation is a direction for future work.

The global climate model simulations might have biases due to coarse resolution ($\mathcal{O}(100\,\mathrm{km})$) and convective parameterization \cite{dias2018,weber2021,ji2025}, while the cloud-resolving model simulations only consider the equatorial Pacific and only sustain gravity waves. Therefore, future work may run global cloud-resolving model simulations to more accurately study the coupling between convection and large-scale waves, and to explore whether the transition to periodic precipitation is gradual or abrupt.





\clearpage 

%
\bibliography{science_template}
\bibliographystyle{sciencemag}

%
%
%
%
%
%


\section*{Acknowledgments}
\paragraph*{Funding:}
The authors acknowledge that they received no funding in support for this research.
\paragraph*{Author contributions:}
HQ, YZ, GD, SF contributed to writing of the manuscript. HQ and YZ conceived the study. HQ performed all numerical simulations. HQ performed the main data analysis.
\paragraph*{Competing interests:}
The authors declare that they have no competing interests.
\paragraph*{Data and materials availability:}
All data needed to evaluate the conclusions in the paper are present in the paper and/or the Supplementary Materials.
\subsection*{Supplementary materials}
Materials and Methods\\
Figs. S1 to S9\\
Table S1\\
References \textit{(35-\arabic{enumiv})}\\ 


\newpage


\renewcommand{\thefigure}{S\arabic{figure}}
\renewcommand{\thetable}{S\arabic{table}}
\renewcommand{\theequation}{S\arabic{equation}}
\renewcommand{\thepage}{S\arabic{page}}
\setcounter{figure}{0}
\setcounter{table}{0}
\setcounter{equation}{0}
\setcounter{page}{1} 


\begin{center}
\section*{Supplementary Materials for\\ \scititle}


Heng~Quan$^{1,2\ast}$,
Yi~Zhang$^{3}$,
Guy~Dagan$^{4}$,
Stephan~Fueglistaler$^{1,2}$\and

\small$^\ast$Corresponding author(s). Email(s): hengquan@princeton.edu

\end{center}

\subsubsection*{This PDF file includes:}
Materials and Methods\\
Figures S1 to S9\\
Tables S1\\

\newpage


\subsection*{Materials and Methods}



\subsubsection*{Cloud-resolving simulations}

We use the System for Atmospheric Modeling (SAM, \cite{khairoutdinov2003}) version 6.11.5 cloud-resolving model (CRM). The model is nonhydrostatic, uses one-moment bulk microphysics and a simple Smagorinsky-type scheme for subgrid turbulence, and computes the surface sensible heat, latent heat and momentum fluxes based on the Monin–Obukhov similarity theory. All simulations use a $\mathrm{CO_2}$ concentration of 355\,ppm, and the concentrations
of other trace gases are set to the default values. The sea surface temperature (SST) is fixed in all simulations.

Most simulations (except those with SST warmer than 310\,K) have a vertical grid of 64 levels, starting at 25\,m and extending up to 27\,km, and the vertical grid spacing increases from 50\,m at the lowest levels to roughly 1\,km at the top of the domain. The model has a rigid lid at the top with a wave-absorbing layer occupying the upper third of the domain to prevent the reflection of gravity waves. The domain size is varied over a wide range of sizes and geometries, and the horizontal resolution is 2\,km.

A time step of 5\,s is used. Radiative fluxes are calculated every 5 min using the CAM (Community Atmosphere Model) radiation scheme \cite{collins2006}, except for experiments using prescribed radiative cooling
rates. Following \cite{lutsko2024}, the incoming solar radiation is fixed at 650.83\,$\mathrm{Wm^{-2}}$. Small temperature perturbations are added near the surface at the beginning of the simulation to initialize convection. All simulations are run for 140 days and reach equilibrium after approximately 50 days. All our results are based on the last 40 days of hourly model output. 

We run two types of CRM simulations - small-domain simulations resolving convection-scale dynamics, and mock Walker simulations \cite{kuang2012} resolving both convection-scale dynamics and its interaction with large-scale dynamics. They are summarized in table \ref{table:SAM_simulations}. 

For small-domain simulations, the domain size is $128\,\mathrm{km} \times 128\,\mathrm{km}$ and we use doubly-periodic boundary conditions. The SST is uniform and ranging from 290\,K to 325\,K with an increment of +5\,K. The vertical grid is composed of 81 levels, which follows the RCEMIP protocol \cite{wing2018RCEMIP} up to a height of 33 km and is extended to 40 km with 1 km resolution to
account for the deeper troposphere under higher SSTs. We have another group of small-domain simulations coupled to gravity waves (\cite{kuang2008DGW}, more details below) with SST ranging from 290\,K to 310\,K with an increment of +5\,K. We repeat the 300\,K simulation with two modifications: one with the large-scale humidity advection induced by gravity waves multiplied by a factor of 2, the other with the temperature advection multiplied by 1.5. The factors 1.5 and 2 are roughly the ratio of the vertical gradient of potential temperature and specific humidity between the 310\,K simulation and the 300\,K simulation as shown in Figure \ref{fig:SAM_WTG_mechanism}. ("2xQGRAD" and "1.5xTGRAD" in table \ref{table:SAM_simulations}, more details below). 

For mock Walker simulations, the domain size is $12288\,\mathrm{km} \times 128\,\mathrm{km}$ for 3-D simulations and $12288\,\mathrm{km}$ for 2-D simulations. We use solid wall boundary conditions at the two edges in the long dimension and periodic boundary conditions in the short dimension (for 3-D runs). The SSTs linearly decrease by 8\,K from the left boundary ($x=0$) to the right boundary ($x = 12288$\,km), mimicking the east-west SST gradient across the equatorial Pacific and causing a overturning circulation. The domain average SST ranges from 290\,K to 310\,K with an increment of +5\,K. We have two more 2-D 310\,K mock Walker simulations with smaller domain size $L/2 = 6144\,\mathrm{km}$ and $L/3 = 4096\,\mathrm{km}$, and they also have a 8\,K SST contrast between two edges. We have another group of 2-D mock Walker simulations with the interactive radiative transfer disabled by prescribing a uniform radiative cooling rate $Q_{\mathrm{rad}} = -1.5\, \mathrm{K\,day^{-1}}$ throughout the troposphere (where the temperature is warmer than $207.5$\,K) and using a Newtonian relaxation towards $200$\,K in the stratosphere ("fixed QRAD" in table \ref{table:SAM_simulations}) \cite{pauluis2006}.

\subsubsection*{Coupling gravity waves to small-domain CRM simulations}

We use the method from \cite{kuang2008DGW} to parameterize gravity waves in small-domain CRM simulations and study the interaction between convection and large-scale dynamics. We start by considering the linearized large-scale wave equations in a non-rotating 2-D anelastic system:

\begin{equation}
\label{equ:momentum}
    \frac{\partial u'}{\partial t} = - \frac{\partial \phi'}{\partial x} - \varepsilon u',
\end{equation}

\begin{equation}
\label{equ:continuity}
    \frac{\partial u'}{\partial x} + \frac{\partial \omega'}{\partial p} = 0,
\end{equation}

\begin{equation}
\label{equ:hydrostatic}
    \frac{\partial \phi'}{\partial p} = - \frac{R_d T'}{p}.
\end{equation}

These equations are the perturbation equations of momentum, continuity and hydrostatic balance. $\varepsilon$ is the momentum damping coefficient and other symbols assume their standard meteorological meaning. The virtual effect is negligible in this system \cite{kuang2008DGW,wong2023}, so we use temperature instead of virtual temperature in equation \ref{equ:hydrostatic}. We assume $u',w',T',\phi'$ have a single horizontal wavenumber $k = \frac{2 \pi}{\lambda}$, so that $\frac{\partial}{\partial x} = ik$. After eliminating $u'$ and $\phi'$ from Eqs.\ref{equ:momentum}-\ref{equ:hydrostatic}, we have

\begin{equation}
    (\frac{\partial}{\partial t} + \varepsilon) \frac{\partial^2 \omega'}{\partial p^2} = k^2 \frac{R_d T'}{p}.
\end{equation}

Because the dimension of the small-domain simulations ($\mathcal{O}(100\,\mathrm{km})$) is much smaller than the wavelength of the gravity waves that we consider ($\mathcal{O}(10000\,\mathrm{km})$, the long dimension of the mock Walker runs), we interpret $w'$ as the spatially-uniform vertical velocity induced by large-scale gravity waves, i.e. $\omega_{\mathrm{large-scale}}$. Similarly, $T'$ is interpreted as the deviation of the domain-average temperature from a reference radiative convective equilibrium (RCE) state, $\langle T \rangle - T_{\mathrm{RCE}}$. Therefore, we have

\begin{equation}
\label{equ:omega}
    \left (\frac{\partial}{\partial t} + \varepsilon \right ) \frac{\partial^2 \omega_{\mathrm{large-scale}}}{\partial p^2} = k^2 \frac{R_d (\langle T \rangle - T_{\mathrm{RCE}})}{p}.
\end{equation}

The vertical velocity due to large-scale gravity waves advects temperature and specific humidity in the small-domain, so the tendency equations of domain-average temperature and specific humidity are written as

\begin{equation}
    \frac{\partial \langle T \rangle}{\partial t} = S_T \underbrace{- \omega_{\mathrm{large-scale}} \left (\frac{p}{p_0} \right)^{\frac{R_d}{c_p}} \frac{\mathrm{d} \theta_{\mathrm{RCE}}}{\mathrm{d} p}}_{\left (\frac{\partial \langle T \rangle}{\partial t} \right)_{\mathrm{large-scale}}},
    \label{equ:T_tendency}
\end{equation}

\begin{equation}
    \frac{\partial \langle q \rangle}{\partial t} = S_q \underbrace{- \omega_{\mathrm{large-scale}} \frac{\mathrm{d} q_{\mathrm{RCE}}}{\mathrm{d} p}}_{\left (\frac{\partial \langle q \rangle}{\partial t} \right)_{\mathrm{large-scale}}}.
    \label{equ:q_tendency}
\end{equation}

The source terms $S_T$ and $S_q$ are convective-scale tendencies simulated explicitly by the model, and we add the terms $\left (\frac{\partial \langle T \rangle}{\partial t} \right)_{\mathrm{large-scale}}$ and $\left (\frac{\partial \langle q \rangle}{\partial t} \right)_{\mathrm{large-scale}}$ to represent the effects of large-scale gravity waves. The resulting $\langle T \rangle$ is used in equation \ref{equ:omega} to diagnose $\omega_{\mathrm{large-scale}}$ for the next time step.

The reference RCE profiles $T_{\mathrm{RCE}}$, $\theta_{\mathrm{RCE}}$ and $q_{\mathrm{RCE}}$ are obtained by averaging the corresponding variables over the last 40 days of the small-domain simulations without gravity wave coupling, and they are the initial conditions of the small-domain simulations coupled to gravity waves. We use $\varepsilon = 0.25\,\mathrm{day}^{-1}$ and $\lambda = 8000\,\mathrm{km}$ in all simulations. A different choice might affect the exact critical SST for the transition to periodic convection, but the transition still occurs as the SST warms (not shown).

\subsubsection*{Global climate model simulations}

We use the Geophysical Fluid Dynamics Laboratory (GFDL) atmospheric general circulation model AM4 \cite{zhao2018a,zhao2018b} to conduct global atmospheric simulations. AM4 uses a horizontal grid spacing of approximately 100\,km and saves the data to disk on a grid with 180 grid points in the meridional, and 288 grid point in the zonal direction (i.e. $1.0^{\circ} \times 1.25^{\circ}$ for the horizontal resolution). The greenhouse gas concentrations and aerosol emissions correspond to the conditions of the year 2000 in all simulations. The control simulation (“present climate”) is forced by the observed climatological (1982-2001) monthly means of SSTs and sea ice concentrations from the HADISST1 dataset \cite{Rayner2003}, and the zonal SST profile in the equatorial Pacific region is similar to the 300\,K mock Walker simulation performed with SAM. We have a uniform SST+5\,K simulation and a uniform SST+10\,K simulation, comparable to the 305\,K and 310\,K mock Walker simulations, respectively. All AM4 simulations are integrated for 15~years, with the first 5~years discarded to eliminate spin-up effects. All our results are based on the last 10 years of 3-hourly model output.

\subsubsection*{Computing the power spectrum of the precipitation time series}

We compute the power spectrum of the 3-hourly precipitation time series in selected regions over the last 10 years of the GCM simulations by a bootstrap method. We first divide the raw series by the time-average precipitation, so that the change of the power spectrum is independent of the change in time-average precipitation. Then we group the 10-year precipitation series into many 50-day chunks. We randomly shuffle the order of those 50-day chunks for 1000 times and get 1000 new precipitation series. These new time series retain the $\mathcal{O}(10\,\mathrm{day})$ signals in the original time series, but no longer have low-frequency signals with periods longer than $\mathcal{O}(50\,\mathrm{day})$. We calculate the power spectrum of each new precipitation series and show the average of the 1000 power spectrums in Fig.\,\ref{fig:AM4_precip_series}g-j. Doing so results in a smoother power spectrum than the spectrum of the original precipitation series and enables a significance test: We find that the $\mathcal{O}(10\,\mathrm{day})$ spectral power of the +10\,K simulation (red curves in Fig.\,\ref{fig:AM4_precip_series}g-j) is larger than the $\mathcal{O}(10\,\mathrm{day})$ spectral power at the 90th percentile among the 1000 spectrums of the ``present climate" simulation in all selected regions (not shown), confirming that the intensification of the $\mathcal{O}(10\,\mathrm{day})$ oscillation is significant at the $p=0.1$ level.

\subsubsection*{Computing the equatorial wave spectrum}

We compute the equatorial ($10^\circ$\,S to $10^\circ$\,N) wave spectrum from the GCM output according to the method in \cite{wheeler1999} and \cite{macdonald2024}. Because linear equatorial waves are either symmetric or antisymmetric about the equator, we first decompose any given variable into a symmetric component $FS(\varphi) = \frac{F(\varphi) + F(-\varphi)}{2}$ and an antisymmetric component $FA(\varphi) = \frac{F(\varphi) - F(-\varphi)}{2}$, where $\varphi$ is the latitude. Then we partition the time series of each component into 192-day segments, overlapping by 128 days. We remove the mean and linear trend from each segment and apply 2-D discrete Fourier transforms in time and zonal directions. We average the spectrum power, i.e. squared modulus of the spectrum, over all segments and all latitudes. To get the background spectrum featuring red noise, we apply a 1-2-1 filter 40 times in the frequency dimension and 10 times in the wavenumber dimension to the raw spectrum. We calculate the significance level by taking the ratio of the raw power spectrum to the smoothed background spectrum. A higher ratio means a stronger wave signal, and the 95\% significance level of the ratio is 1.232 according to a chi-squared test.





\begin{figure}
    \centering
        \includegraphics[width=\linewidth]{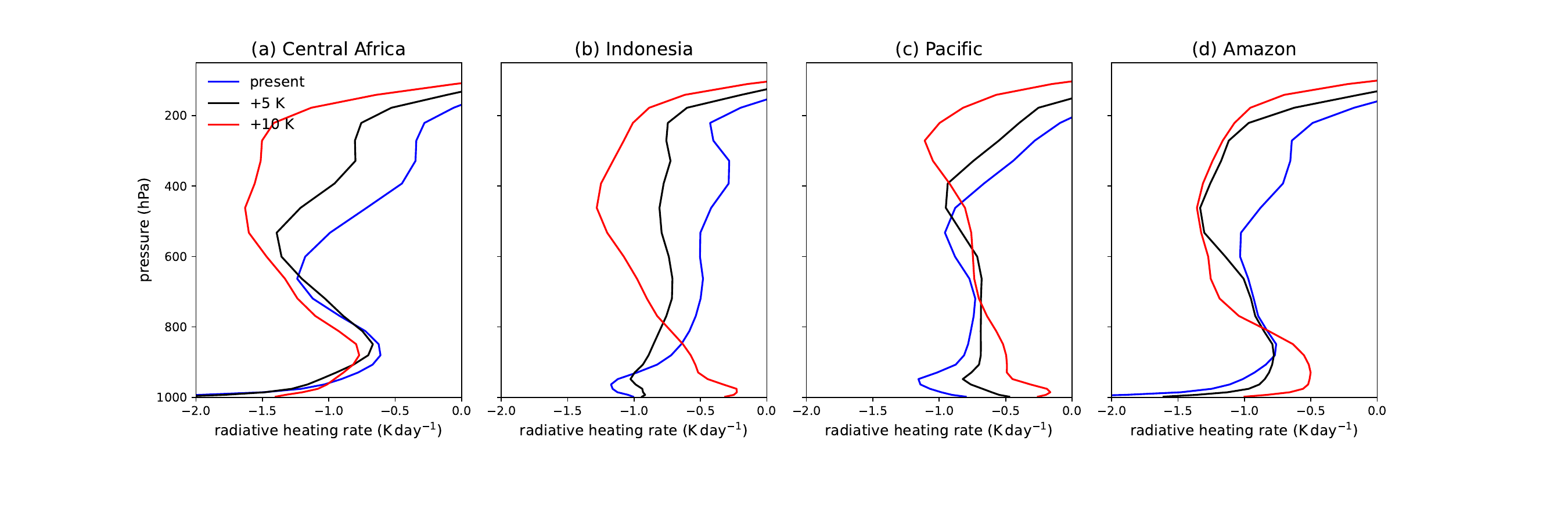}
    \caption{\textbf{The transition to periodic precipitation in global climate model simulations is not due to large vertical gradient in radiative cooling rate.} Results here are the vertical profiles of time-average radiative heating rates in the periods and regions as shown in Figure \ref{fig:AM4_precip_series}.}
    \label{fig:AM4_QRAD_tmean}
\end{figure}

\begin{figure}
    \centering
        \includegraphics[width=\linewidth]{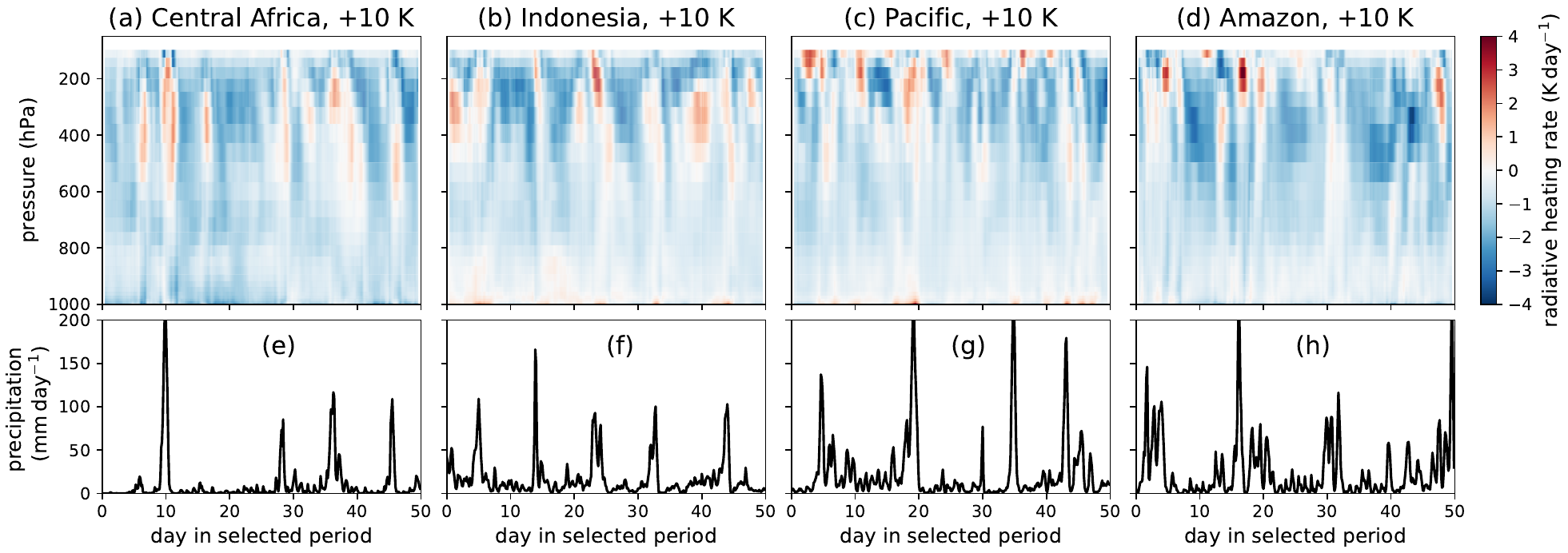}
    \caption{\textbf{The transition to periodic precipitation in global climate model simulations is not due to large vertical gradient in radiative cooling rate.} 
    \textbf{a-d}, The time evolution of the radiative cooling rate in the ``+10\,K" global climate model simulation in the periods and regions as shown in Figure \ref{fig:AM4_precip_series}.
    \textbf{e-h}, The corresponding time evolution of precipitation.}
    \label{fig:AM4_QRAD_series}
\end{figure}

\begin{figure}
    \centering
    \includegraphics[width=\linewidth]{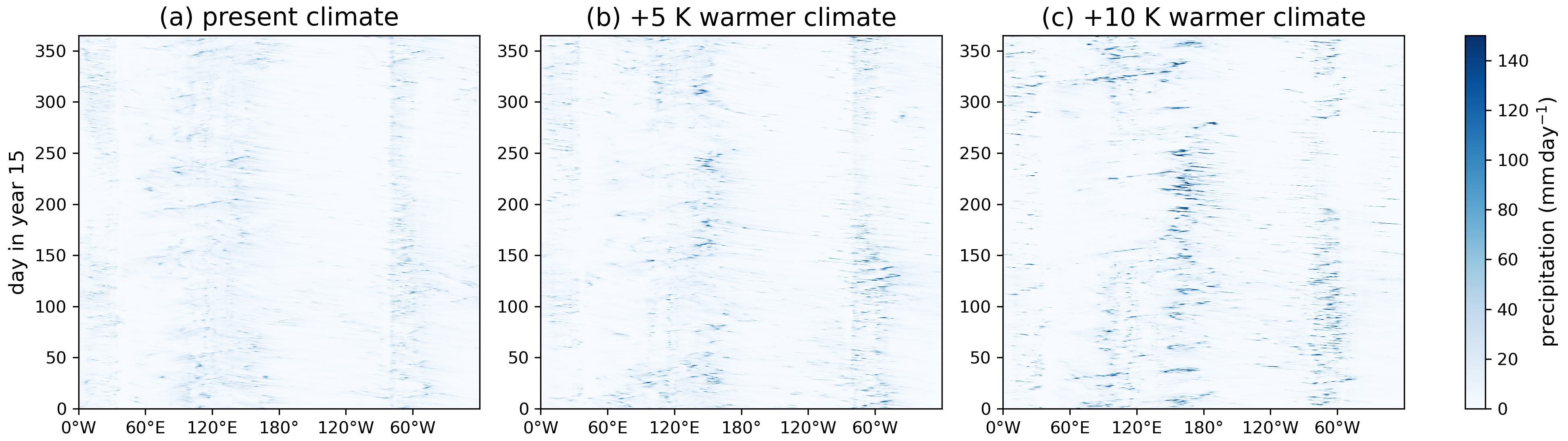}
    \caption{\textbf{The intensification of convectively-coupled Kelvin waves under global warming in global climate model simulations.} \textbf{a-c}, The Hovmoller diagrams of equatorial precipitation ($5^\circ$S-$5^\circ$N average) for present climate, a +5\,K warmer climate and a +10\,K warmer climate.}
    \label{fig:AM4_precip_Hovmoller}
\end{figure}

\begin{figure}
    \centering
    \includegraphics[width=\linewidth]{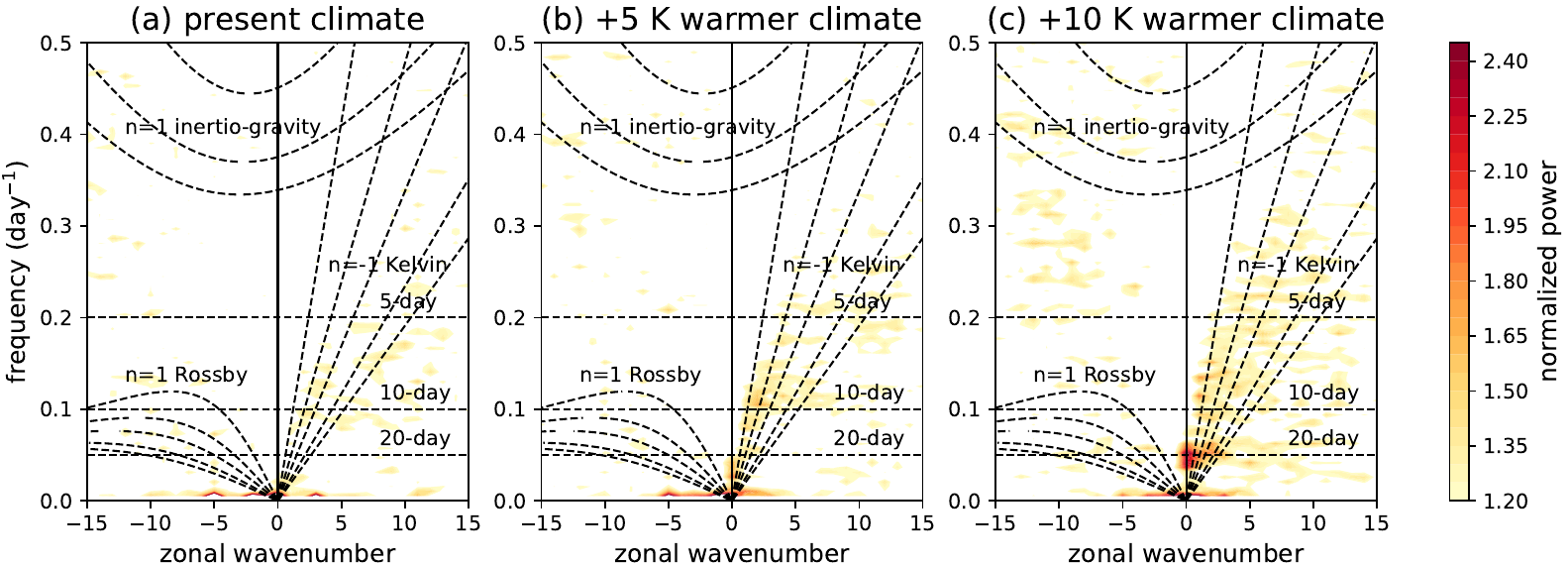}
    \caption{\textbf{The intensification of convectively-coupled Kelvin waves under global warming in global climate model simulations.} \textbf{a-c}, The space–time spectral diagrams (see Materials and Methods) of the spectral power of the symmetric component of equatorial precipitation for present climate, a +5\,K warmer climate and a +10\,K warmer climate (the spectral power is averaged over $10^\circ$S-$10^\circ$N before computing the background spectrum). The significance threshold is 1.232 based on a chi-squared test. The dashed lines show dispersion curves for equatorial Kelvin waves (with meridional index -1), Rossby waves (with meridional index 1) and inertio-gravity waves (with meridional index 1) with equivalent depths of 8, 12, 25, 50, and 100\,m.}
    \label{fig:AM4_EQwaves_symmetric}
\end{figure}

\begin{figure}
    \centering
        \includegraphics[width=0.5\linewidth]{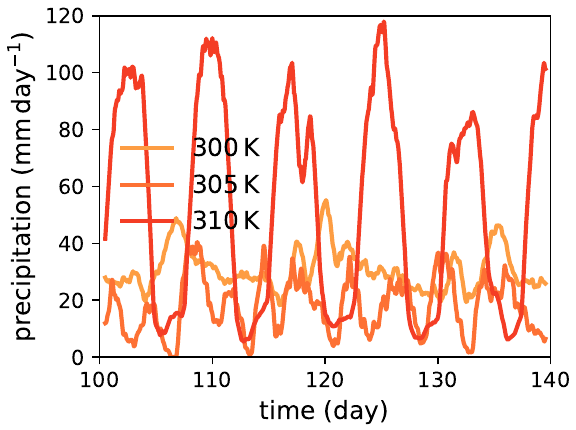}
    \caption{\textbf{The transition to periodic precipitation in mock Walker simulations can occur without lower-tropospheric radiative heating.} Same as Figure \ref{fig:SAM_precip_series}\textbf{c}, but for 2-D mock Walker simulations with a prescribed and uniform radiative cooling rate $Q_{\mathrm{rad}} = -1.5\,\mathrm{K\,day^{-1}}$ throughout the troposphere.}
    \label{fig:SAM_precip_series_mock_Walker_uniform_QRAD}
\end{figure}

\begin{figure}
    \centering
    \includegraphics[width=\linewidth]{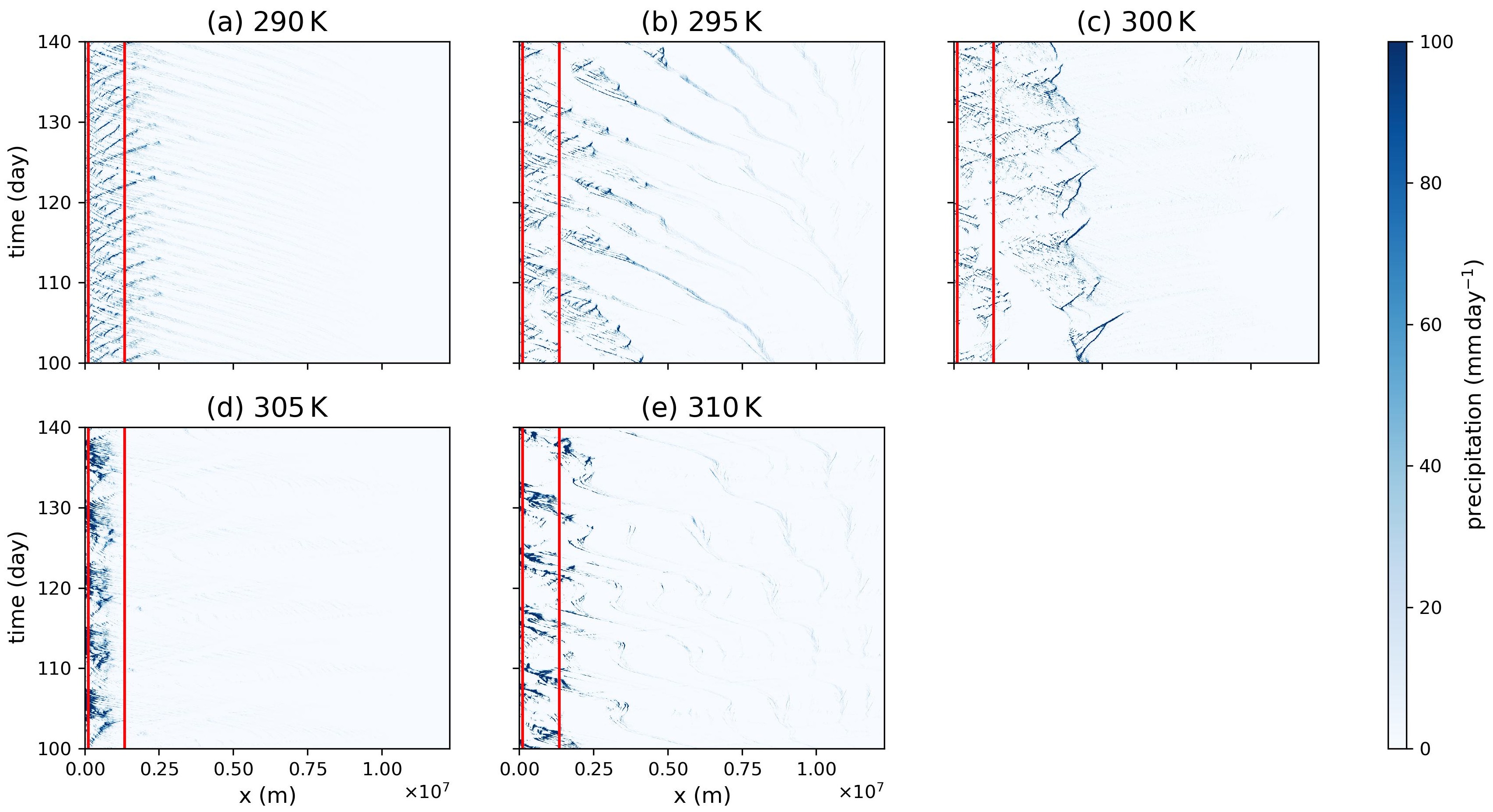}
    \caption{\textbf{The intensification of convectively-coupled gravity waves under global warming in mock Walker simulations.} \textbf{a-e}, The Hovmoller diagrams of precipitation (averaged along the short dimension) for 3-D mock Walker simulations.}
    \label{fig:SAM_mock_Walker_Hovmoller}
\end{figure}

\begin{figure}
    \centering
    \includegraphics[width=\linewidth]{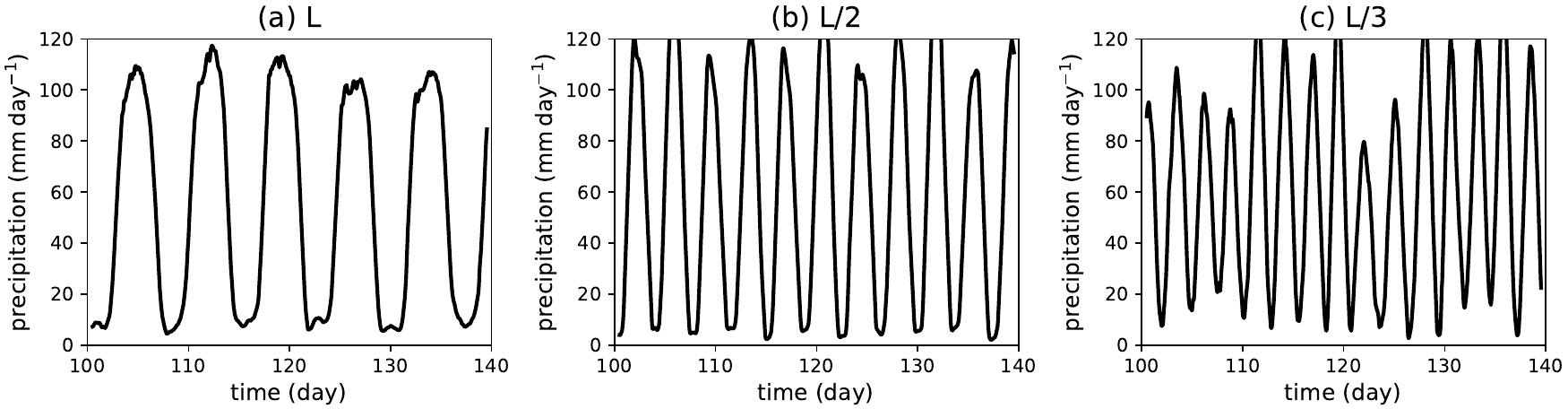}
    \caption{\textbf{The period of precipitation is equal to the period of gravity waves and proportional to the domain size.} Same as Figure \ref{fig:SAM_precip_series}\textbf{c}, but for the 2-D 310\,K mock Walker simulation with different domain sizes.}
    \label{fig:SAM_precip_series_mock_Walker_perturbL}
\end{figure}

\begin{figure}
    \centering
        \includegraphics[width=0.5\linewidth]{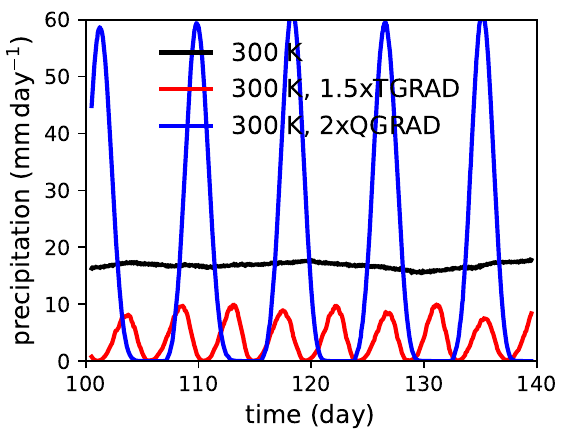}
    \caption{\textbf{Larger vertical gradients of potential temperature or specific humidity alone can cause the transition to periodic precipitation.} The black line is the 40-day domain-mean precipitation time series in the small-domain simulation coupled to gravity waves with SST at 300\,K. The red line is from the same simulation, but with the wave-induced temperature advection ($\left (\frac{\partial \langle T \rangle}{\partial t} \right)_{\mathrm{large-scale}}$ in equation \ref{equ:T_tendency}) multiplied by 1.5. The blue line is also from the same simulation, but with the wave-induced humidity advection ($\left (\frac{\partial \langle q \rangle}{\partial t} \right)_{\mathrm{large-scale}}$ in equation \ref{equ:q_tendency}) multiplied by 2. They are equivalent to multiplying the vertical $\theta$ and $q$ gradients when calculating the wave-induced advective tendencies. The factors 1.5 and 2 are roughly the ratio of the vertical gradient of potential temperature and specific humidity between the 310\,K simulation and the 300\,K simulation as shown in Figure \ref{fig:SAM_WTG_mechanism}. }
    \label{fig:SAM_precip_series_WTG_+Tqgrad}
\end{figure}


\begin{figure}
    \centering
    \includegraphics[width=0.75\linewidth]{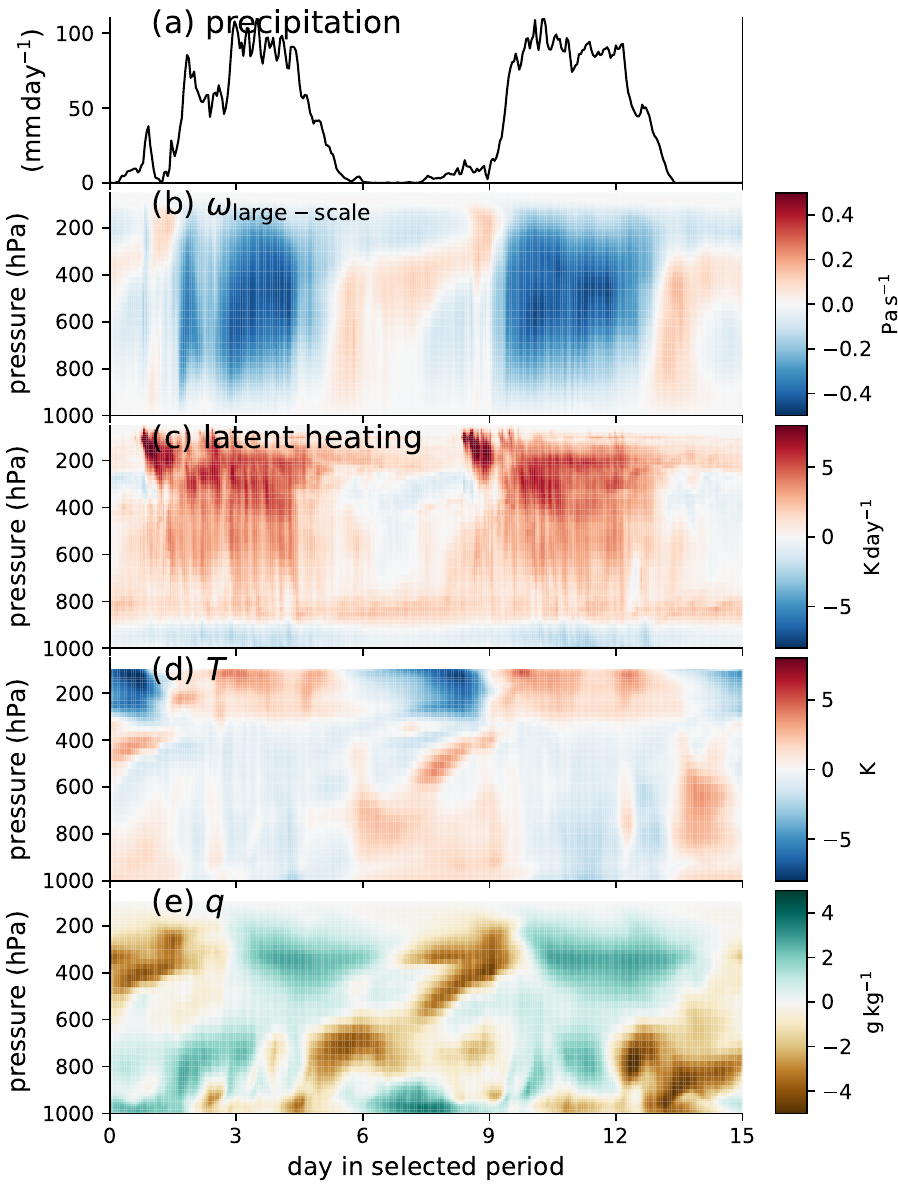}
    \caption{\textbf{The precipitation cycles in a warmer climate.} Same as Figure \ref{fig:convection_cycle_comparison}, but for the 310\,K mock Walker simulation. All panels except \textbf{c} show averages over the top 10\% SST, and \textbf{c} shows the domain-mean latent heating rate.}
    \label{fig:SAM_mock_Walker_convection_cycle}
\end{figure}



\begin{table}
\caption{List of cloud-resolving simulations performed with SAM}
\label{table:SAM_simulations}
\begin{tabular}{@{}llll@{}}
\hline
simulation & SST\,(K) & radiation & domain size \\
\hline
small-domain & 290 to 325 in 5 intervals & interactive & $128\,\mathrm{km} \times 128\,\mathrm{km}$ \\
small-domain + gravity wave & 290, 295, 300, 305, 310 & interactive & $128\,\mathrm{km} \times 128\,\mathrm{km}$ \\
small-domain + gravity wave, 2xQGRAD & 300 & interactive & $128\,\mathrm{km} \times 128\,\mathrm{km}$ \\
small-domain + gravity wave, 1.5xTGRAD & 300 & interactive & $128\,\mathrm{km} \times 128\,\mathrm{km}$ \\
mock Walker 3D & 290, 295, 300, 305, 310 & interactive & $12288\,\mathrm{km} \times 128\,\mathrm{km}$ \\
mock Walker 2D & 290, 295, 300, 305, 310 & interactive & $12288\,\mathrm{km}$ \\
mock Walker 2D, L/2 & 310 & interactive & $6144\,\mathrm{km}$ \\
mock Walker 2D, L/3 & 310 & interactive & $4096\,\mathrm{km}$ \\
mock Walker 2D, fixed QRAD & 290, 295, 300, 305, 310 & fixed & $12288\,\mathrm{km}$ \\
\hline
\end{tabular}
\end{table}




\end{document}